# Real-valued parametric conditioning of an RNN for interactive sound synthesis

**Lonce Wyse**
Communications and New Media Department
National University of Singapore
Singapore
lonce.acad@zwhome.org

**Abstract**

A Recurrent Neural Network (RNN) for audio synthesis is trained by augmenting the audio input with information about signal characteristics such as pitch, amplitude, and instrument. The result after training is an audio synthesizer that is played like a musical instrument with the desired musical characteristics provided as continuous parametric control. The focus of this paper is on conditioning data-driven synthesis models with real-valued parameters, and in particular, on the ability of the system a) to generalize and b) to be responsive to parameter values and sequences not seen during training.

## Introduction

Creating synthesizers that model sound sources is a laborious and time consuming process that involves capturing the complexities of physical sounding bodies or abstract processes in software and/or circuits. For example, it is not enough to capture the acoustics of a single piano note to model a piano because the timbral characteristics change in nonlinear ways with both the particular note struck and the force with which it is struck. Sound modeling also involves capturing or designing some kind of interface that maps input control signals such as physical gestures to sonic qualities. For example, clarinets have keys for controlling the effective length of a conically bored tube, and a single-reed mouthpiece that is articulated with the lips, tongue, and breath, all of which effect the resulting sound. Writing down the equations and implementing models of these processes in software or hardware has been an ongoing challenge for researchers and commercial manufacturers for many decades.

In recent years, deep learning neural networks have been used for data-driven modeling across a wide variety of domains. They have proven adept at learning for themselves what features of the input data are relevant for achieving their specified tasks. "End-to-end" training relieves the need to manually engineer every stage of the system and generally results in improved performance.

For sound modeling, we would like the system to learn the association between parametric control values provided as input and target sound as output. The model must generate a continuous stream of audio (in the form of a sequence of sound samples), responding with minimal delay to continuous parametric control. A recurrent neural network (RNN) is developed herein since the sequence-oriented architecture is an excellent fit for an interactive sound synthesizer. During training of the RNN, input consists of audio "augmented" with parameter values, and the system learns to predict the next audio sample conditioned on the input audio and parameters. The input parameters consist of musical pitch, volume, and an instrument identifier, and the target output consists of a sequence of samples comprising a musical instrument tone characterized by the three input parameters.

The focus of this paper is not on the details of the architecture, but on designing and training the control interface for sound synthesizers. Various strategies for conditioning generative RNNs using augmented input have been developed previously under a variety of names including "side information," "auxiliary features," and "context" (Mikolov & Zweig, 2012; Hoang, Cohn, and Haffari, 2016). For example, phonemes and letters are frequently used for conditioning the output of speech systems. However, phonemes and letters are discrete and nominal (unordered) while the control parameters for synthesizers are typically ordered and continuously valued. Some previous research has mentioned conditioning with pitch, but real-valued conditioning parameters for generative control have not received much attention in experiments or documentation.

In this paper, the following questions will be addressed: If a continuously valued parameter is chosen as an interface, then how densely must the parameter space be sampled during training? How reasonable (for the sound modeling task) is the synthesis output during the generative phase using control parameter values not seen during training? Is it adequate to train models on unchanging parametric configurations, or must training include every sequential combination of parameter values that will be used during synthesis? How responsive is the system to continuous and discrete (sudden) changes to parameter values during synthesis?

## Previous Work

Mapping gestures to sound has long been at the heart of sound and musical interface design. Fels and Hinton (1993) described a neural network for mapping hand gestures to parameters of a speech synthesizer. Fiebrink (2011) developed the Wekinator for mapping arbitrary gestures to parameters of sound synthesis algorithms. Fried and Fiebrink (2013) used stacked autoencoders for reducing the dimensionality of physical gestures, images, and audio clips, and then used the compressed representations to map between domains. Françoise et al. (2014) developed a "mapping-by-demonstration" approach taking gestures to parameters of synthesizers. Fasciani and Wyse (2012) used machine learning to map vocal gestures to sound and separately to map from sound to synthesizer parameters for generating sound. Gabrielli et al. (2017) used a convolutional neural network to learn upwards of 50 "microparameters" of a physical model of a pipe organ. However, all of the techniques described above use predefined synthesis systems for sound generation, and are thus limited by the capabilities of the available synthesis algorithms. They do not support the learning of mappings between gestures and arbitrary sound sequences that would constitute "end to end" learning including the synthesis algorithms themselves.

Recent advances in neural networks hold the promise of learning end-to-end models from data. WaveNet (Van den Oord et al., 2016) is a convolutional network, and SampleRNN (Mehri et al., 2016) is a recurrent neural network that both learn to predict the "next" sample in a stream conditioned on what we will refer to as a "recency" window of preceding samples. Both can be conditioned with external input supplementing the sample window to influence sound generation. For example, a coded representation of phonemes can be presented along with audio samples during training in order to generate desired kinds of sounds during synthesis.

Engel et al. (2017) address parametric control of audio generation for musical instrument modeling. They trained an autoencoder on instrument tones, and then used the activations in the low-dimensional layer connecting the encoder to the decoder as sequential parametric "embedding" codes for the instrument tones. Each instrument is thus represented as temporal sequence of low-dimensional vectors. The temporal embeddings learned in the autoencoder network are then used to augment audio input for training the convolutional WaveNet (Van den Oord et al., 2016) network to predict audio sequences. During synthesis, it is possible to interpolate between the time-varying augmented vector sequences representing different instruments in order to generate novel instrument tones under user control.

The current work is also aimed at data-driven learning of musical instrument synthesis with interactive control over pitch and timbre. It differs from Engel et al. in that all learning and synthesis is done with a single network, and the network is a sequential RNN, small, and oriented specifically to study properties of continuous parameter conditioning relevant for sound synthesis.

## Architecture

The synthesizer is trained as an RNN that predicts one audio sample at the output for each audio sample at the input (Figure 1). Parameter values for pitch, volume, and instrument are concatenated with the input and presented to the system as a vector with four real-valued components normalized to the range [0,1].

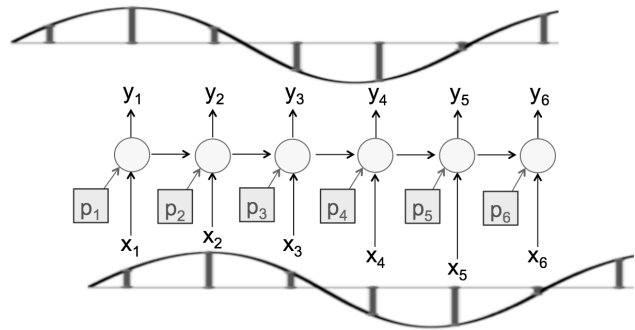

*Figure 1. The RNN unfolded in time. During training, audio (x) is presented one sample per time step with the following sample as output. The conditioning parameters associated with the data such as pitch (p) are concatenated with the audio sample as input. During generation, the output at each time step (e.g. y1) becomes the input (e.g. x2) at the next time step, while the parameters are provided at each time step by the user.*

To manage the length of the sequences used for training, a sampling rate of 16kHz for audio is used which, with a Nyquist frequency of 8kHz, is adequate to capture the pitch and timbral features of the instruments and note ranges used for training. Audio samples are mu-law encoded which provides a more effective resolution/dynamic range trade-off than linear coding. Each sample is thus coded as one of 256 different values, and then normalized to provide the audio input component. The target values for training are represented as one-hot vectors, with each node representing one of the 256 possible sample values.

The network consists of a linear input layer mapping the four-component input vector (audio, pitch, volume, and instrument) to the hidden layer size of 40. This is followed by a 4-layer RNN with 40 gated recurrent unit (GRU) (Cho et al., 2014) nodes each and feedback from each hidden layer to itself. A final linear layer maps the deepest GRU layer activations to the one-hot audio output representation (see Figure 2). An Adam optimizer (Kingma and Ba, 2015) was used for training, with weight changes driven by cross-entropy error and the standard backpropagation through time algorithm (Werbos, 1995). Uniform noise was added at 10% of the volume scaling for each sequence, and no additional regularization (drop-out, normalization) techniques were used. During generation, the maximum-valued output sample is chosen, mu-law encoded, and then fed back as input for the next time step.

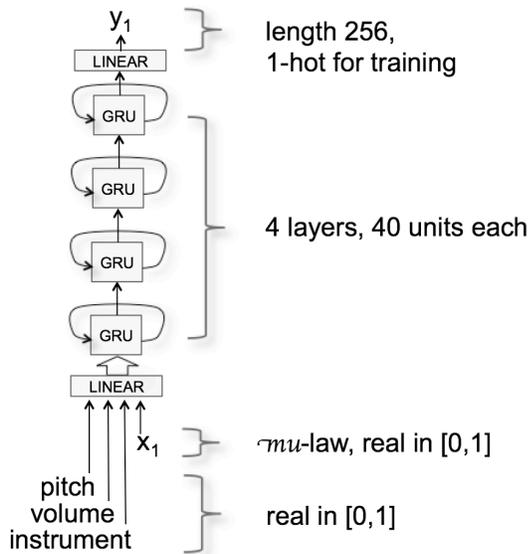

*Figure 2. The network consists of 4 layers of 40 GRU units each. A four-dimensional vector is passed through a linear layer as input and the output is a one-hot encoded audio sample.*

For training data, two synthetic and two natural musical instruments were used (see Table 1). For the synthetic instruments, one was comprised of a fundamental frequency at the nominal pitch value and even numbered harmonics (multiples of the fundamental), and the other comprised of the fundamental and odd harmonics.

The two natural instruments are a trumpet and a clarinet from the NSynth database (Engel et al., 2017). Thirteen single recordings of notes in a one-octave note range (E4 to E5) were used for each of the instruments for training (see Figure 3). "Steady state" audio segments were extracted from the NSynth files by removing the onset (0-.5 seconds) and decay (3-4 seconds) segments from the original recordings. The sounds were then normalized so that all had the same root-mean-square (rms) value. Labels for the pitch parameters used for input were taken from the NSynth database (one for each note, despite any natural variation in the recording), while different volume levels for training were generated by multiplicatively scaling the sounds and taking the scaling values as the training parameter. Sequences of length 256 were then randomly drawn from these files for training. At the 16kHz sample rate, 256 samples covers 5 periods of the fundamental frequency of the lowest pitch used. Sequences were trained in batches of 256.

| 1. Synth even | |
|---|---|
| 2. Synth odd | |
| 3. Trumpet | |
| 4. Clarinet | |

*Table 1. Waveform samples for the four instruments used for training on the note E4 (fundamental frequency ~ 330). The first two instruments are synthetically generated with even and odd harmonics respectively; the Trumpet and Clarinet are recordings of physical instruments from the NSynth database.*

### Pitch and the learning task

Musical tones have a "pitch" which is identified with a fundamental frequency. However, pitch is a perceptual phenomenon, and physical vibrations are rarely exactly periodic. Instead, pitch is perceived despite a rich variety of different types of signals and "noise." Even the sequence of digital samples that represent the synthetic tones do not generally have a period equal to their nominal pitch value unless the frequency components of the signal happen to be exact integer submultiples of the sampling rate.

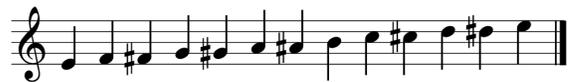

*Figure 3. A chromatic scale of 13 notes spanning one octave, E4 (with a fundamental frequency of ~333 Hz) to E5 (~660 Hz) used for training the network.*

The goal of training is to create a system that synthesizes sound with the pitch, volume, and instrumental quality that are provided as parametric input during generation. However, the system is not trained explicitly to produce a target pitch, but rather to produce single samples conditioned on pitch (and other) parameter values and a recency window of audio samples. Since the perception of pitch is established over a large number of samples (at least on the order of the number of samples in a pitch period), the network will have the task of learning distributions of samples at each time step, and must learn to depend on long-term dependencies to prevent pitch "errors" from accumulating.

## Generalization

For synthesizer usability, we require that continuous control parameters map to continuous acoustic characteristics. This implies the need for generalization in the space of the conditioning parameters. For example, the pitch parameter is continuously valued, but if training is conducted only on a discrete set of pitch values, we desire that during generation, interpolated parameter values produce pitches that are interpolated between the trained pitch values. This is similar to what is expected in regression tasks (except that regression outputs are explicitly trained, and sound model pitch is only implicitly trained, as discussed above).

**Training: Synthetic instrument, pitch endpoints only**

In order to address the question of how densely the real-valued musical parameter spaces, particularly pitch, must be sampled, the network was first trained with synthetically generated tones with pitches only at the two extreme ends of the scale for the training data and parameter range.

After training only the endpoints, the generative phase is tested with parametric input. Figure 4 shows a spectrogram of the synthesizer output, as the pitch parameter is swept linearly across its range of values from its lowest to highest and back. The pitch is smoothly interpolated across the entire range of untrained values. The output is clearly not linear in the parameter value space. Rather, there is a "sticky" bias in the direction of the trained pitches, and a faster than linear transition in between the extreme parameter values. Also visible is a transition region half way between the trained values where the synthesized sound is not as clear (visibly and auditorily) as it is at the extremities. This interpolation behavior is perfectly acceptable for the goal of synthesizer design.

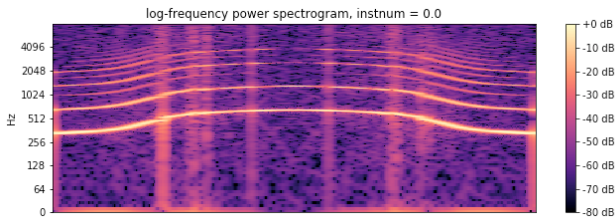

*Figure 4. A network was trained only on the extreme low and high pitches at the endpoints of the one-octave parameter value range. During generation, the parameter value was swept through untrained values between its lowest and its highest and back again over 3 seconds. The result is this continuously, although non-linearly varying pitch.*

## Responsiveness

Another feature required of an interactive musical sound synthesizer is that it must quickly respond to control parameter changes so that they have immediate effect on the output produced. We would like to be free of any constraints on parameter changes (e.g. smoothness). Thus the question arises as to whether the system will have to be trained on all possible sequential parameter value combinations in order to respond appropriately to such sequences during synthesis. It would consume far less time to train on individual pitches than on every sequential pitch combination that might be encountered during synthesis. However, this would mean that at any time step where a parameter is changed during synthesis, the system would be confronted not only with an input configuration not seen during training, but with a parameter value representing a pitch in "conflict" with the pitch of the audio in the recency window responsible for the current network activation.

To explore this question of responsiveness, the model was trained only on individual pitches. Then for the generative phase, it was presented with a parameter sequence of notes values spaced out over the parameter range, specifically an E-major chord (E4, G#4, B4, E5) played forward and backward as a 7-note sequence over a total duration of 5 seconds.

As can be seen in Figure 5, the system was able to respond to the parameter values to make the desired changes to sample sequences for the new pitches. The pitches produces in response to each particular parameter value are the same as those produced during the sweep through the same values.

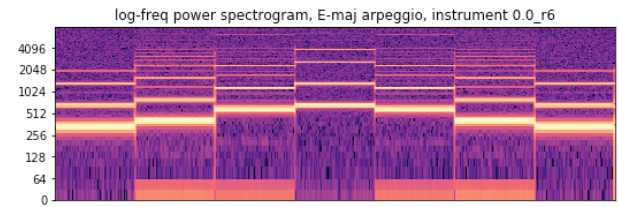

*Figure 5. The trained "Synth even" instrument controlled with an arpeggio over the pitch range illustrates the model's ability to respond quickly to pitch changes. This image also shows that the untrained middle pitch values are not synthesized as clearly as the trained values at the extremities. Furthermore, the middle values contain both even and odd harmonics, thus combining timbre from each of the two trained instruments.*

It can also be seen in Figure 5 that the response to untrained pitch parameters are less clear than those at the extreme. They are also richer in harmonics, including some of the odd harmonics present only in the other trained instrument (*Synth odd*). There is also a non-zero transition time between notes indicated by the vertical lines visible in the spectrogram. They have a duration of approximately 10ms and actually add to the realistic quality of the transition.

A related issue to responsiveness is drift. Previous work (Glover, 2015) trained networks to generate musical pitch, but controlled the pitch production at the start of the generative phase by "priming" the network with trained data at the desired pitch. However, the small errors in each sample accumulate in this kind of autoregressive model, so that the result is a sound that drifts in pitch. When using the augmented input described here which supplies continuous information about the desired pitch to the network, there was never any evidence of drifting pitch. For the same reason that new pitch parameter values "override" the audio history as the control parameter changes in the sweep and the arpeggio, the pitch parameter plays the role of preventing drift away from its specified value.

## Physical instrument data
### Training: Natural instruments, pitch endpoints only
When the system was trained on real data from the trumpet and clarinet recordings in the NSynth database, the pitch interpolation under the 2-pitch extreme endpoint training condition was less pronounced than for the synthetic instrument data. The smooth but nonlinear pitch sweep was present for the trumpet, but for the clarinet, the "stickiness" of the trained values extended almost across the entire untrained region, making a fast transition in the middle between the trained values (Figure 6). One potential explanation for this contrasting behavior is that real instruments exhibit quite different waveforms at different pitches, while for the synthetic data, the waveform was exactly the same at all pitches, changing only in frequency with correspondingly less demanding interpolation requirements.

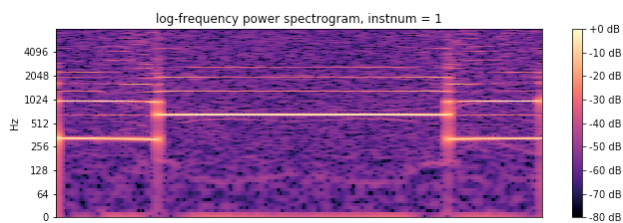

*Figure 6. When the network was trained on real data with only two extreme values of pitch, pitch had a more pronounced "stickiness" to extreme trained pitch values showing a transition region without a glide as the pitch parameter moves smoothly from low to high and back.*

The "stickiness" bias toward trained pitches is also quite acceptable for synthesizers driven with sparse data in the parameter space. However, the 2-endpoint pitch training regimen was far more extreme than the sampling that would be typical for synthesizer training. In fact, when the system is trained on notes in the chromatic scale (each note spaced in frequency from its neighbor by approximately 6%), the interpolation of pitch is still seen for physical instrument data (see below).

Knowing the tendency of the system to generate interpolated pitch output in response to untrained conditioning parameter values, and knowing that is not necessary to train on combinatorial parameter sequences in order to get responsiveness to parameter changes during the generative phase, we can now be confident about choosing a training regimen for musical instrument models.

### Training: Natural instruments, 13 pitches per octave
When this RNN model is trained with 2 instruments, 24 volume levels, and a 13-note chromatic scale across an octave, thereby augmenting the audio sample stream with 3 real-valued conditioning parameters, then the behavior of the trained model is what we would expect from a musical instrument synthesizer. Stable and accurate pitches are produced for trained parameter values, and interpolated pitches with the proper instrument timbre are produced for in-between values (Figure 7a). The system is immediately sensitive and responsive to parameter changes (Figure 7b), and as the instrument parameter changes smoothly across the untrained space between the two trained endpoints, the timbre changes while holding pitch and volume fairly stable (Figure 7c).

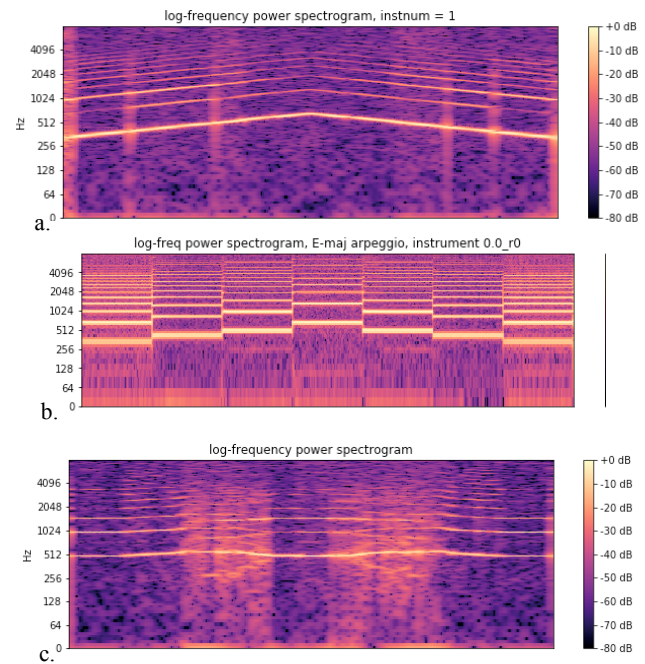

*Figure 7. a. The clarinet, trained on 13 chromatic notes across an octave, generating a sweep with the pitch parameter swept from low to high and back. b. The trumpet playing the arpeggio pattern. c. A continuous forth-and-back sweep across the instrument parameter trained with the natural trumpet and clarinet at its endpoint values.*

## Future Work

Several directions are suggested for future work. Range and sound quality will have to be improved for the system to be a performance instrument. Extending the pitch range beyond one octave, and in particular to notes in lower registers, would require more training and a network capable of learning longer time dependencies, especially if a higher sampling rate were used to improve quality. The architecture would seem to lend itself to unpitched sound "textures" that vary in perceptual dimensions other than pitch, as well. However, based on preliminary experiments, training will be more difficult than for the semi-periodic pitched sounds explored here, and interpolation in dimensions such as "roughness" seem more challenging than pitch. Finally, the synthesis phase, even with modest size of the current system, is still slower than real time. However, given the one sample in / one sample out architecture, and with only a few layers in between, there are no in-principle obstacles to a low latency system so important for musical performance.

## Conclusions

An RNN was trained to function as a musical sound synthesizer capable of responding continuously to real-valued control values for pitch, volume, and instrument type. The audio input sequence data was augmented with the desired parameters to be used for control during synthesis.

Key usability characteristics for generative synthesizers were shown to hold for the trained RNN model: the ability to produce reasonable pitch output for untrained parameter values, and the ability to respond quickly and appropriately to parameter changes. The training data can be quite sparse in the space defined by the conditioning parameters, and still generate sample sequences appropriate for musical sound synthesis. We also showed that a classic "drifting" pitch problem is addressed with the augmented input strategy, even though pitch is only implicitly trained in this autoregressive audio sample prediction model. This bodes will for the use of RNNs for developing general data-driven sound synthesis models.

## Supplementary Media

Audio referenced in this paper, as well as links to open-source code for reproducing this data can be found online at: http://lonce.org/research/RNNAudioConditioning/

## Acknowledgements

This research was supported in part by an NVidia Academic Programs GPU grant.